\documentclass[aps,pre,twocolumn,superscriptaddress,tight]{revtex4-2}

\usepackage{amssymb}
\usepackage{amsfonts}
\usepackage{color}
\usepackage{graphicx}
\usepackage{amsmath}
\usepackage{times}
\usepackage{bm}
\usepackage{float}
\usepackage{lipsum}
\usepackage{import}
\usepackage{enumerate}
\usepackage[colorlinks,linkcolor=blue,anchorcolor=blue,citecolor=blue]{hyperref}

\begin{document}

\title{Inferring bifurcation diagrams of two distinct chaotic systems by a single machine}

\author{Jianmin Guo}
\affiliation{School of Physics and Information Technology, Shaanxi Normal University, Xi'an 710062, China}

\affiliation{College of Physics and Electronic Information Engineering, Qinghai Normal University, Xining 810008, China}

\author{Yao Du}
\affiliation{School of Physics and Information Technology, Shaanxi Normal University, Xi'an 710062, China}

\author{Yizhen Yu}
\affiliation{School of Physics and Information Technology, Shaanxi Normal University, Xi'an 710062, China}

\author{Yong Zou}
\affiliation{School of Physics and Electronic Science, East China Normal University, Shanghai 200062, China}

\author{Xingang Wang}
\email{Corresponding author: wangxg@snnu.edu.cn}
\affiliation{School of Physics and Information Technology, Shaanxi Normal University, Xi'an 710062, China}

\begin{abstract}

We propose a dual-channel reservoir-computing scheme for inferring the dynamics of two distinct chaotic systems with a single machine. By augmenting a standard reservoir with a system-label channel and a parameter-control channel, the machine can be trained from time series collected from a few sampled states of the two systems. We show that the trained machine not only predicts the short-time evolution of the sampled states, but also reproduces the long-term statistical properties of unseen states, thereby enabling reconstruction of the bifurcation diagrams of both systems from partial observations. The effectiveness of the scheme is demonstrated for the Lorenz and R\"ossler systems in numerical simulations and for the Chua and R\"ossler circuits in experiments. Functional-network analysis further shows that the two target systems are encoded by distinct dynamical patterns in the reservoir. These results extend multifunctional and parameter-aware reservoir computing, and provide a route to data-driven inference of multiple nonlinear systems using a single machine.

\end{abstract}

\date{\today}
\maketitle

\section{Introduction}

Model-free, data-driven inference of chaotic systems using reservoir computing (RC) has attracted considerable interest in recent years~\cite{RC:Maass2002,RC:Jaeger,RC:Pathak2017,RC:LZ2018,RC:Fan,RC:digtwin,RC:DY2024,RC:adaptiveRCLYC,RC:NC2024perspective}. In RC, a fixed recurrent network maps the input data into a high-dimensional dynamical representation, and the target dynamics are learned by training only the readout matrix. Compared with many deep-learning architectures, RC has a particularly simple training procedure: it contains only a single recurrent layer, and all parameters except those in the output layer remain fixed after construction. This feature makes RC computationally efficient and attractive for hardware implementation~\cite{RC:Tanaka2019}. Despite its simple architecture, RC has shown outstanding performance in a wide range of data-driven tasks~\cite{RC:adaptiveRCLYC,RC:NC2024perspective,RC:Tanaka2019}, especially in the prediction of chaotic dynamics. For typical chaotic systems, RC can predict the state evolution accurately for about $10$ Lyapunov times~\cite{RC:Jaeger,RC:Pathak2017,RC:LZ2018}, substantially extending the prediction horizon of traditional approaches in nonlinear science. Beyond short-term prediction, RC can also reproduce the long-term statistical properties of chaotic dynamics, i.e., the ``climate"~\cite{RC:Pathak2017}. More recently, parallel RC schemes have been developed for learning spatiotemporal dynamics~\cite{RC:Pathak2018,RC:Parlitz2018,RC:ParallelMachinePRL2022}, opening a route toward applications of RC to real-world complex systems.

While early studies of RC focused mainly on learning the dynamics of a specific system, recent efforts have moved toward more general frameworks that can either infer unseen dynamics or emulate multiple systems of distinct dynamics with a single machine~\cite{RC:CK2020,RC:Guo2021,KLW:2021,RC:Kim2021,RC:FHW2021,RC:LHB2024,RC:HBL2025,RC:LZX2020,RC:Flynn2021,KLW2024,RC:MFRC2025}. By introducing a parameter-control channel into the input layer, parameter-aware RC (PARC) has been developed to infer dynamics not encountered during training~\cite{KLW:2021,RC:Kim2021}. In PARC, the training data consist of time series collected from several states of the same system, together with their associated bifurcation parameters. During inference, by specifying only the bifurcation parameter, the machine can reproduce the dynamics of both sampled states and unseen states outside the training set. PARC has been successfully applied to a variety of tasks, including prediction of critical transitions~\cite{KLW:2021}, anticipation of synchronization~\cite{RC:FHW2021}, reconstruction of bifurcation diagrams of chaotic circuits~\cite{RC:LHB2024}, and sustaining collective dynamics in complex networks~\cite{RC:HBL2025}. A key limitation of PARC, however, is that the training and inferred states must belong to the same underlying system, i.e., the machine remains monofunctional.

Multifunctionality, by contrast, refers to the ability of a single system to perform different tasks. In neuroscience, multifunctionality is a hallmark of brain dynamics and is essential for diverse cognitive functions~\cite{MF:PAG}. For example, the brain can store many memories and retrieve a specific one without confusion~\cite{Memory:RMS,Memory:RC}. Motivated by the memory function of the brain, different schemes of multifunctional RC (MFRC) have been proposed for learning chaotic systems with distinct dynamics~\cite{RC:LZX2020,RC:Flynn2021,KLW2024,RC:MFRC2025}. Depending on how the target dynamics are stored and retrieved, existing MFRC schemes can be broadly classified into two categories: content-addressable memory (CAM) and location-addressable memory (LAM)~\cite{CAM:1992,Memory:LN}. In CAM-based MFRC, the machine architecture is essentially the same as that of standard RC, and the target dynamics are retrieved using partial dynamical information from the desired system, such as a short time series. Representative examples include the scheme based on invertible generalized synchronization~\cite{RC:LZX2020} and the one based on blended training data~\cite{RC:Flynn2021}. In LAM-based MFRC, a label channel is introduced into the input layer, and a specific dynamics can be retrieved by inputting only the system label~\cite{KLW2024,RC:MFRC2025}. Compared with CAM-based MFRC, LAM-based MFRC is more convenient, since no dynamical content is required for retrieval, and potentially more scalable when many systems are stored~\cite{KLW2024}, making it attractive for practical applications.

Although LAM-based MFRC can, in principle, emulate multiple systems with distinct dynamics, its present performance remains unsatisfactory. Two limitations are particularly important. First, the retrieval success rate is low~\cite{KLW2024}. From the perspective of dynamical systems, multifunctionality requires the reservoir to be multistable~\cite{MF:NP}: depending on the input label and the initial condition, the reservoir may converge to different attractors. Because the corresponding basins of attraction are typically intricate and often fractal, even small perturbations in the initial condition can drive the reservoir to an unintended attractor, leading to retrieval failure. This difficulty becomes more severe as the number of stored systems increases. Second, the retrieved dynamics are often seriously distorted~\cite{KLW2024,RC:MFRC2025}. Unlike standard RC, which can faithfully reproduce the statistical properties of the target dynamics, existing MFRC schemes tend to preserve only coarse qualitative features, such as the number of scrolls and the approximate ranges of the state variables, while failing to reproduce the detailed geometry and statistics of the ground-truth attractors. These limitations substantially restrict the applicability of LAM-based MFRC and motivate the development of new schemes with improved reliability and accuracy.

The main objective of the present work is to develop a new MFRC scheme that can infer the dynamics of multiple chaotic systems both reliably and accurately. To this end, we augment a standard RC with two additional input channels: one labels the target system and the other specifies the bifurcation parameter. We show that, under the joint guidance of the system label and bifurcation parameter, a single machine can not only predict the short-term evolution of each system with high accuracy, but also reproduce the long-term statistical properties of unseen states. Building on this capability, we further demonstrate that, using time series collected from only a few sampled states of each system, the machine can reconstruct the full bifurcation diagrams of both systems with high fidelity. In this way, the proposed scheme combines multifunctionality and parameter awareness within a single RC framework. Our results provide a new route toward high-performance MFRC and a step toward its application to real-world systems.

The rest of the paper is organized as follows. In Sec.~II, we introduce the dual-channel RC scheme. In Sec.~III, we demonstrate its performance on the Lorenz and R\"ossler systems. In Sec.~IV, we analyze its working mechanism from the perspective of functional networks. In Sec.~V, we present the experimental results for chaotic circuits. Finally, in Sec.~VI, we summarize the main results and discuss possible directions for future work.

\section{Method}\label{sec:method}

Our dual-channel RC scheme combines the ideas of PARC and MFRC by incorporating two additional channels into the input layer of the standard RC. Specifically, the machine consists of three modules: the input layer, the reservoir network, and the output layer. The reservoir network and output layer are the same as those of the standard RC, whereas the input layer contains three channels: one for the state vector of the target system, one for the bifurcation parameter, and one for the system label. The input layer is characterized by the matrix $W_{\rm in}\in\mathbb{R}^{D_r\times D_{\rm in}}$, where $D_r$ is the number of nodes in the reservoir network and $D_{\rm in}$ is the dimension of the input vector $\tilde{u}$. The elements of $W_{\rm in}$ are drawn randomly from a uniform distribution over $[-\xi,\xi]$. The role of the input layer is to embed the low-dimensional input vector into the high-dimensional phase space of the reservoir dynamics. The input vector has the form
\begin{equation}
\tilde{u}_{\beta}(t)=[u(t);\beta(t);\bm{l}]^{T},
\end{equation}
where $u(t)\in\mathbb{R}^{D_s}$ denotes the state vector of the target system, $\beta(t)$ is a scalar representing the bifurcation parameter associated with the state $u(t)$, and $\bm{l}\in\mathbb{R}^{2}$ is a binary vector labeling the system, i.e., $[1,0]^T$ for System 1 and $[0,1]^T$ for System 2. For simplicity, we consider only systems with the same value of $D_s$, such as the Lorenz and R\"ossler systems studied below. As illustrated in Fig.~1(a), the training data are formed by concatenating the time series collected from $m$ sampled states of System 1 and $n$ sampled states of System 2. The bifurcation parameters of the sampled states are denoted by $\beta_{1,i}$ ($i=1,\ldots,m$) for System 1 and $\beta_{2,j}$ ($j=1,\ldots,n$) for System 2. As shown in Fig.~1(a), $\beta(t)$ is a stepwise function of time, whereas the label vector remains constant for time series collected from the same system.

The input vector is coupled to the reservoir through the matrix $W_{\rm in}$. In this way, the state vector, bifurcation parameter, and labeling vector are injected into the reservoir through their corresponding channels in the input layer. The reservoir is a recurrent network containing $D_r$ nonlinear nodes. The connectivity among the nodes is defined by a sparse random adjacency matrix $A\in\mathbb{R}^{D_r\times D_r}$, constructed as an Erd\H{o}s-R\'enyi network: each connection has a probability $p$ of having a nonzero weight, and the nonzero weights are drawn from a uniform distribution over $[-1,1]$. The matrix is then rescaled so that its spectral radius is $\lambda$. The initial states of the reservoir nodes are chosen randomly from the interval $[-1,1]$, and the reservoir state, $\bm{r}(t)\in\mathbb{R}^{D_r}$, is updated according to
\begin{equation}
\bm{r}(t+\Delta t)=(1-\alpha)\bm{r}(t)+\alpha\tanh\left[A\bm{r}(t)+W_{\rm in}\tilde{u}_{\beta}(t)\right],
\end{equation}
where $\Delta t$ is the time step for updating the reservoir, $\alpha\in(0,1]$ is the leaking rate, and $\tanh(\cdot)$ is applied componentwise.

\begin{figure*}[tbp]
\begin{center}
\includegraphics[width=0.74\linewidth]{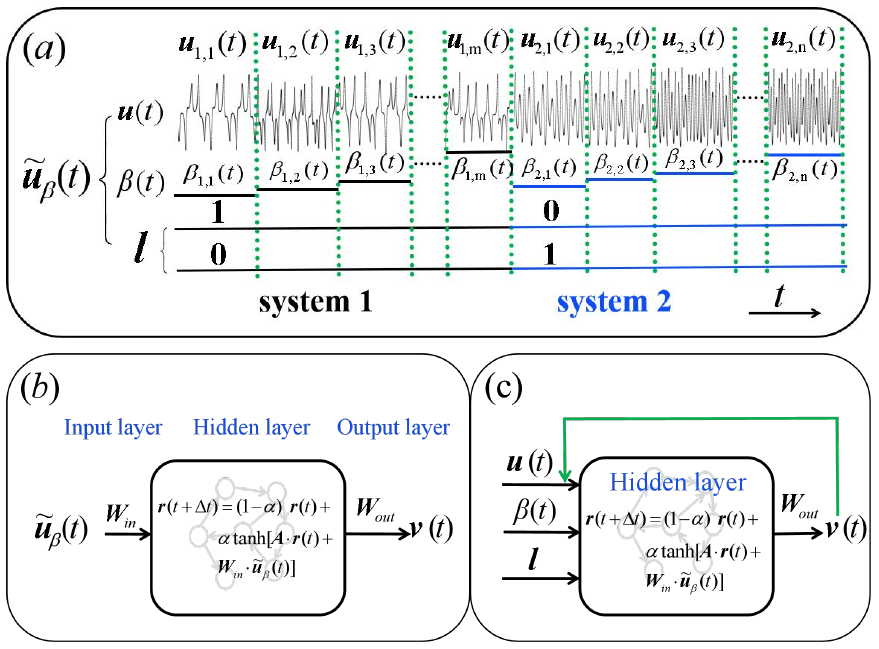}
\centering
\caption{Schematic illustration of the dual-channel RC scheme. (a) Structure of the training data, formed by concatenating the time series collected from sampled states of the two target systems, together with their associated bifurcation parameters and system labels. (b) Open-loop operation in the training phase, where the reservoir is driven by the input data to determine the output matrix. (c) Closed-loop operation in the validation and application phases, where the predicted output is fed back into the reservoir, while the bifurcation parameter and system label are externally specified.}
\vspace{0.18cm}
\label{fig1}
\end{center}
\end{figure*}

The function of the output layer is to generate the output vector $\bm{v}(t+\Delta t)\in\mathbb{R}^{D_{\rm out}}$ (with $D_{\rm out}=D_s$) from the instantaneous state of the reservoir,
\begin{equation}
\bm{v}(t+\Delta t)=W_{\rm out}\tilde{\bm{r}}(t+\Delta t),
\end{equation}
where $W_{\rm out}\in\mathbb{R}^{D_{\rm out}\times(2D_r+1)}$ is the output matrix, and $\tilde{\bm{r}}(t)\in\mathbb{R}^{2D_r+1}$ is the augmented state vector defined by
\begin{equation}
\tilde{r}_1=1,
\end{equation}
\begin{equation}
\tilde{r}_i=r_i,\quad \mbox{for } i=2,\ldots,D_r+1,
\end{equation}
\begin{equation}
\tilde{r}_j=r_j^2,\quad \mbox{for } j=D_r+2,\ldots,2D_r+1.
\end{equation}
Except for the output matrix, all other settings of the machine are fixed once the machine has been constructed.

The implementation of the machine consists of three phases: training, validation, and application. The goal of the training phase is to determine the output matrix from the training data. In this phase, the reservoir is operated in the open-loop mode shown in Fig.~1(b). To estimate the output matrix, the reservoir is first updated for a transient period of $\tau$ steps to eliminate the influence of the initial conditions. The output matrix $W_{\rm out}$ is then calculated as
\begin{equation}
W_{\rm out}=UV^{T}(VV^{T}+\eta I)^{-1},
\end{equation}
where $V\in\mathbb{R}^{(2D_r+1)\times N}$ is the state matrix whose $k$th column is $\tilde{\bm{r}}(k\Delta t)$, $U\in\mathbb{R}^{D_{\rm out}\times N}$ is the target matrix whose $k$th column is $\bm{u}(k\Delta t)$, $I$ is the identity matrix, and $\eta$ is the ridge-regression parameter introduced to avoid overfitting. Here, $N$ denotes the length of the training sequence, i.e., the total number of data points in the training data. The columns of $V$ and $U$ are randomly shuffled before the regression in order to remove temporal correlations among the data points.

The goal of the validation phase is to find the optimal set of hyperparameters for which the machine performs well on both the training and validation datasets. The hyperparameters to be optimized include $D_r$ (the reservoir size), $\xi$ (the range defining the input matrix), $p$ (the connection density of the reservoir), $\lambda$ (the spectral radius of the matrix $A$), $\alpha$ (the leaking rate), and $\eta$ (the regression parameter). In our study, the optimal hyperparameters are obtained numerically by scanning the parameter space using conventional optimization algorithms, e.g., Bayesian optimization. For each set of hyperparameters, the machine performance is evaluated by the root-mean-square error (RMSE) between the predicted time series and the ground truth over the validation data, and the RMSE is averaged over all sampled states. The validation data have the same structure as the training data but a different length. A total of $\tilde{N}$ trials are performed in searching for the optimal hyperparameters. The resulting optimal hyperparameters, together with the associated output matrix, define the optimal machine, which will be used in the subsequent applications.

In the application phase, the machine is operated in the closed-loop mode shown in Fig.~1(c). The tasks of this phase include: (1) predicting the short-term evolution of the sampled states, (2) inferring the long-term statistical properties of unseen states not included in the sampled set, and (3) reconstructing the bifurcation diagrams of both systems from partial observations. (Details of this phase will be given in the following section.)

\section{Model Results}

To demonstrate the efficacy of the proposed dual-channel RC scheme, we apply it to two classical chaotic systems: the Lorenz system and the R\"ossler system. The dynamics of the Lorenz system are governed by the set of equations
\begin{equation}
\begin{aligned}
\dot{x}&=a(y-x),\\
\dot{y}&=rx-xz-y,\\
\dot{z}&=xy-bz,
\end{aligned}
\end{equation}
where the parameters are fixed at $a=10$ and $r=28$, while $b$ is treated as the bifurcation parameter. The dynamics of the R\"ossler system are described by
\begin{equation}
\begin{aligned}
\dot{x}&=-5y-5z,\\
\dot{y}&=5x+2.5y,\\
\dot{z}&=10+5z(x-\sigma),
\end{aligned}
\end{equation}
where $\sigma$ serves as the bifurcation parameter. In the simulations, the initial conditions of both systems are chosen randomly from the interval $(-1,1)$, and the equations are integrated numerically using the fourth-order Runge-Kutta method with time step $\delta t=1\times 10^{-3}$.

For the Lorenz system, the sampled states are chosen as $b\in\{0.2,0.44,0.68,0.92,1.16,1.4\}$ ($m=6$). The system dynamics are periodic for $b\in\{0.2,0.44,0.68\}$ and chaotic for $b\in\{0.92,1.16,1.4\}$. For the R\"ossler system, the sampled states are chosen at $\sigma\in\{3.1,3.175,3.25,3.325,3.4\}$ ($n=5$). The system dynamics are periodic for $\sigma\in\{3.1,3.175\}$ and chaotic for $\sigma\in\{3.25,3.325,3.4\}$. For each sampled state, after discarding a transient of $1\times 10^3$ time steps, we record a time series of the state vector, $u(t)=[x,y,z]^T$, containing $10\,000$ data points with sampling interval $\Delta t=10\delta t$. The data are normalized to the range $[0,1]$. Each data point is then augmented by its associated bifurcation parameter $\beta$ ($b$ for the Lorenz system and $\sigma$ for the R\"ossler system) and label vector $\bm{l}$ ($[1,0]^T$ for the Lorenz system and $[0,1]^T$ for the R\"ossler system) to form the input vector $\tilde{u}_{\beta}(t)$. Each time series is divided into two segments. The first segment contains $6\,000$ data points and is used as the training series for estimating the output matrix. The remaining data points are used as the validation series for optimizing the machine. The training (validation) dataset is constructed by concatenating the training (validation) series from all sampled states ($11$ in total). Consequently, the training and validation datasets contain $66\,000$ and $44\,000$ data points, respectively.

Having prepared the datasets, we next estimate the output matrix and optimize the machine hyperparameters according to the procedure introduced in Sec.~II. In estimating the output matrix, the warm-up series, which is used to eliminate the influence of the reservoir initial conditions, contains $\tau=50$ data points. In optimizing the machine hyperparameters, we fix the reservoir size as $D_r=1000$, while optimizing the hyperparameters $(\xi,p,\lambda,\alpha,\eta)$. The optimal hyperparameters are obtained after $\tilde{N}=400$ trials in parameter space using the \texttt{optimoptions} function in MATLAB. The search ranges of the hyperparameters are $\xi\in[0,3]$, $p\in[0,1]$, $\lambda\in[0,2]$, $\alpha\in[0,1]$, and $\eta\in[10^{-10},10^{-2}]$. The validation phase yields the optimal hyperparameters $(\xi,p,\lambda,\alpha,\eta)=(1.4224,0.6530,1\times 10^{-5},0.3667,9.2\times10^{-3})$.

\begin{figure*}[tbp]
\begin{center}
\includegraphics[width=0.95\linewidth]{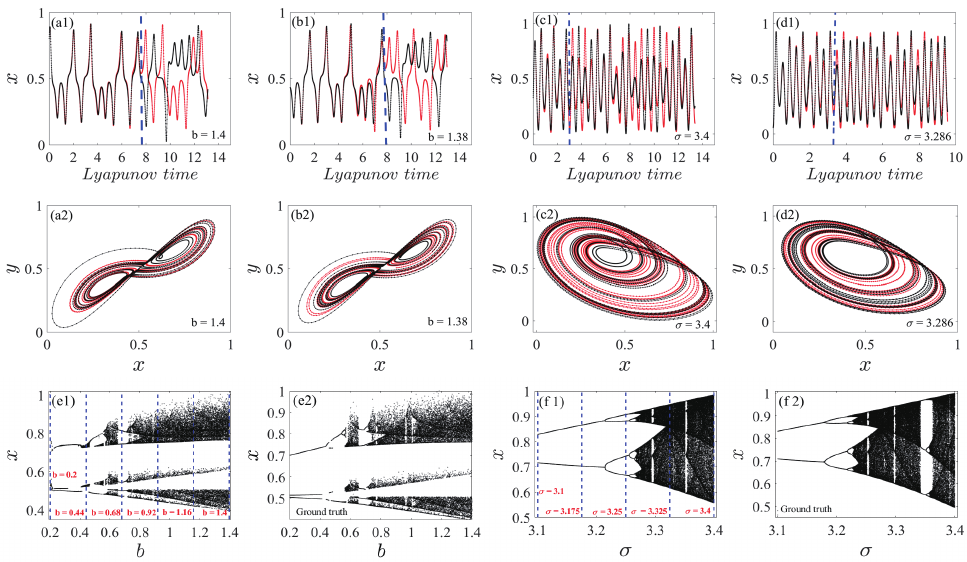}
\centering
\caption{Inferring the dynamics of the Lorenz and R\"ossler systems by dual-channel RC. (a,b,e) Results for the Lorenz system. (a1,a2) State evolution and chaotic attractor for the sampled state $b=1.4$. (b1,b2) State evolution and chaotic attractor for the unseen state $b=1.38$. (e1) Bifurcation diagram inferred by the machine from partial observations. (e2) Ground-truth bifurcation diagram obtained from model simulations. (c,d,f) Results for the R\"ossler system. (c1,c2) State evolution and chaotic attractor for the sampled state $\sigma=3.4$. (d1,d2) State evolution and chaotic attractor for the unseen state $\sigma=3.286$. (f1) Bifurcation diagram inferred by the machine. (f2) Ground-truth bifurcation diagram obtained from model simulations. In (a1)--(d2), black curves denote the results from model simulations and red curves denote the results inferred by the machine; the vertical dashed lines mark the prediction horizons. In (e1) and (f1), the vertical dashed lines indicate the sampled states from which the training data are collected.}
\label{fig2}
\end{center}
\vspace{0.2cm}
\end{figure*}

Will the dynamics of both systems be inferred reliably and accurately by the new scheme? To address this question, we first evaluate the performance of the machine in inferring the dynamics of the Lorenz system. The results for the sampled state $b=1.4$ are shown in Fig.~2(a1). We see that the system state can be predicted accurately for about $8$ Lyapunov times, where the largest Lyapunov exponent of this state is approximately $0.67$. Based on the machine outputs, we plot in Fig.~2(a2) the long-term trajectory of the system state in the $(x,y)$ plane. It is seen that the ``climate'' of the sampled state, i.e., the statistical properties of the chaotic Lorenz attractor, is faithfully reproduced. To quantify the similarity between the inferred attractor and the ground truth, we calculate the deviation value, $D_v$, between them by the method introduced in Ref.~\cite{Noise:SR2023}. The result shows that $D_v\approx0.35$, indicating good overlap in the phase space (two attractors are regarded as well overlapped when $D_v<0.4$). Setting $b=1.38$ as the bifurcation parameter, which is not included in the training data, we show in Figs.~2(b1) and (b2) the inferred state evolution and attractor, respectively. It is found that, for this unseen state, the prediction horizon remains about $8$ Lyapunov times (for this state, the largest Lyapunov exponent is approximately $0.665$), and that the ``climate'' is also well reproduced ($D_v=0.32$). We note that, in generating the attractors shown in Figs.~2(a2) and (b2), the machine is operated in the closed-loop mode [see Fig.~1(c)] throughout the process, i.e., it is driven only by the bifurcation parameter and the label vector.

We next use the machine to reconstruct the bifurcation diagram, starting with the Lorenz system. To this end, we fix the input of the label channel as $\bm{l}=[1,0]^T$ and increase the parameter-control input $\beta=b$ from $0.2$ to $1.4$ with an increment of $1\times 10^{-3}$. As in Fig.~2(b2), the machine is operated in the closed-loop mode, and data are collected from the outputs after a short transient of $50$ iterations. For each value of $b$, we run the machine for $5\times10^{3}$ steps and then extract the local maxima of the $x$ variable from the outputs. The inferred bifurcation diagram is plotted in Fig.~2(e1), and the ground truth obtained from model simulations is shown in Fig.~2(e2). We see that the reconstructed bifurcation diagram captures the main features of the ground truth, including the period-doubling cascades, the periodic windows, and the boundaries of the chaotic regimes. These results in Figs.~2(a), (b), and (e) demonstrate that the dynamics of the Lorenz system are successfully inferred.

We next evaluate the performance of the machine in inferring the dynamics of the R\"ossler system. The results for the sampled state $\sigma=3.4$ are shown in Figs.~2(c1) and (c2). We see that, although the ``climate'' of the system dynamics is faithfully reproduced [as shown in Fig.~2(c2), which has $D_v\approx0.28$], the state evolution can be predicted accurately for only about $3$ Lyapunov times [as shown in Fig.~2(c1)]. Similar results are obtained for the unseen state $\sigma=3.286$, which is not included in the training data, as shown in Figs.~2(d1) and (d2). For this unseen state, the inferred attractor also agrees well with the ground truth, with $D_v\approx0.29$. By increasing $\sigma$ from $3.1$ to $3.4$ with an increment of $1\times 10^{-4}$, we reconstruct the bifurcation diagram from the machine outputs, as shown in Fig.~2(f1). As in the Lorenz case, the bifurcation diagram is plotted using the local maxima of the $x$ variable. The ground-truth bifurcation diagram obtained from model simulations is shown in Fig.~2(f2). We see that the inferred bifurcation diagram is in good agreement with the ground truth. These results in Figs.~2(c), (d), and (f), demonstrate that the dynamics of the R\"ossler system are successfully inferred too.

It is worth emphasizing that the ``climate'' of both systems can be reliably reproduced by the proposed scheme. Additional simulations show that, provided the transient period before data collection is sufficiently long ($>100$ iterations), the same bifurcation diagram can be reconstructed even when the reservoir is initialized with random conditions. Meanwhile, compared with conventional RC schemes designed for learning a single dynamical system, the predictive performance of the present machine is degraded. For example, Figs.~2(c) and (d) show that the prediction horizon of the R\"ossler system is only about $3$ Lyapunov times, which is significantly shorter than that of conventional RC schemes, for which the prediction horizon is typically longer than $6$ Lyapunov times. Although such degraded performance is expected for MFRC schemes~\cite{KLW2024,RC:MFRC2025}, the underlying mechanism remains unclear. In the following section, we investigate the working mechanism of the proposed RC scheme from the perspective of functional networks in neuroscience.

\section{Mechanism Analysis}\label{sec:functional_network}

Functional networks provide an efficient framework for characterizing the correlation structure among coupled dynamical units in complex systems~\cite{FN:CR2005,FN:EB2009,FN:HSY2023}. Unlike physical networks, in which links represent actual connections, the links in functional networks are virtual. Moreover, depending on how the correlations among units are defined, different functional networks can in general be constructed from the same structural network~\cite{FN:ZCS2006,FN:LMH2010,FN:LWJ2015}. Functional-network analysis has proved useful in exploring the functionality of a variety of real-world systems, e.g., the cognitive functions of the human brain~\cite{FN:BB1995,FN:KJF2011,FN:EVM2005}. As an emulator of biological neural systems, RC shares many features with neuronal networks, making functional networks a natural tool for investigating its working mechanism~\cite{RC:LZX2020,RC:MFRC2025,RC:WL2022}. In Ref.~\cite{RC:LZX2020}, it was shown that, in MFRC, different target dynamical systems are represented by distinct functional networks in the reservoir. In Ref.~\cite{RC:MFRC2025}, it was further shown that the performance of MFRC depends crucially on the distinguishability of the functional networks associated with the target systems. Given the improved performance of our dual-channel RC scheme, which not only reproduces the ``climate'' of the sampled states accurately, but also faithfully infers the bifurcation diagrams of both systems, two natural questions arise: (1) Is the improved performance in attractor retrieval associated with enhanced distinguishability between the corresponding functional networks? (2) Are the functional networks of sampled and unseen states distinguishable from each other?

We adopt the method introduced in Ref.~\cite{Arenas2006} to construct functional networks from the reservoir dynamics. To obtain the functional network of the Lorenz (R\"ossler) system, we operate the machine in the closed-loop mode shown in Fig.~1(c), with the system-label channel set to $\bm{l}=[1,0]^T$ ($[0,1]^T$) and the parameter-control channel set to $b$ ($\sigma$). The pairwise correlation between nodes $i$ and $j$ in the reservoir is defined as~\cite{Arenas2006}
\begin{equation}
\rho_{ij}(T)=\langle \cos[r_i(T)-r_j(T)]\rangle,
\label{cor}
\end{equation}
which is evaluated at time $T$ of the reservoir evolution. In Eq.~(\ref{cor}), the angle brackets denote the ensemble average over $N_a$ sets of initial conditions for the same reservoir network topology. According to Ref.~\cite{RC:LZX2020}, accurate replication of the target dynamics requires the reservoir network to synchronize with the target in the generalized sense, namely through the mechanism of invertible generalized synchronization. Once synchronization is achieved, the reservoir dynamics become independent of the initial conditions, and the correlation coefficients defined by Eq.~(\ref{cor}) approach steady values after a transient ($T\gg1$). Setting the threshold coefficient to $\rho_c$, we establish a functional link between nodes $i$ and $j$ whenever $\rho_{ij}>\rho_c$. In this way, we obtain a binary correlation matrix $B$, with $b_{ij}=b_{ji}=1$ if $\rho_{ij}>\rho_c$ and $b_{ij}=0$ otherwise. Treating $B$ as the adjacency matrix, the functional network is then constructed by connecting the reservoir nodes accordingly.

\begin{figure}[tbp]
\begin{center}
\includegraphics[width=\linewidth]{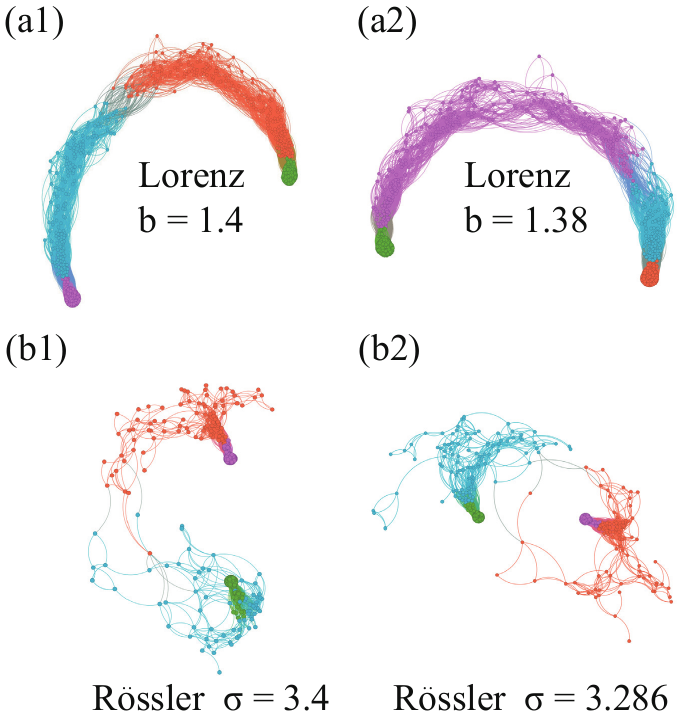}
\centering
\caption{Functional networks extracted from the reservoir for different inferred states. (a1) Lorenz system with $b=1.4$, corresponding to one of the sampled states. (a2) Lorenz system with $b=1.38$, corresponding to an unseen state not included in the training set. (b1) R\"ossler system with $\sigma=3.4$, corresponding to one of the sampled states. (b2) R\"ossler system with $\sigma=3.286$, corresponding to an unseen state not included in the training set.}
\label{fig3}
\end{center}
\end{figure}

The functional network of the Lorenz attractor with parameter $b=1.4$, corresponding to one of the sampled states, is shown in Fig.~3(a1). The parameters used for constructing the functional network are $N_a=200$, $T=1000$, and $\rho_c=0.95$ (qualitatively similar results are obtained when these parameters are varied slightly). We see that the functional network is composed of distinct communities~\cite{CN:Newman}, with nodes within each community densely connected, whereas connections between different communities are sparse. Numerical analysis based on the method introduced in Refs.~\cite{Modularity:VDB,Modularity:RL} shows that the network consists of four communities, with the network modularity being approximately $0.55$. Since the reservoir network is constructed by connecting the nodes randomly, these communities emerge as a self-organized response of the reservoir to the input data. The functional network of the Lorenz system with parameter $b=1.38$, which is not included in the training phase, is shown in Fig.~3(a2). This network also consists of four communities, and its modularity is about $0.55$ too.

The functional networks of the R\"ossler system with parameters $\sigma=3.4$ and $\sigma=3.286$, corresponding to one sampled state and one unseen state, are shown in Figs.~3(b1) and (b2), respectively. Both networks consist of four communities. For the sampled state shown in Fig.~3(b1), the network modularity is about $0.503$; for the unseen state shown in Fig.~3(b2), the network modularity is about $0.504$.

Upon an initial inspection of the functional networks shown in Fig.~\ref{fig3}, we find that the networks associated with states from the same system are similar, whereas those associated with states from different systems are distinct. To quantify the similarity and difference between the functional networks, we calculate the normalized mutual information (NMI) between them using the method introduced in Refs.~\cite{Newman2004,Karrer2008,Meil2007}. The calculation of NMI involves two steps. The first is to partition each functional network into communities, which is done using the algorithm introduced in Refs.~\cite{Modularity:VDB,Modularity:RL}. The second is to calculate the mutual information between the resulting community partitions,
\begin{equation}
I_{\mathrm{NMI}}(\mathbf{C}_1,\mathbf{C}_2)=\frac{2I(\mathbf{C}_1,\mathbf{C}_2)}{H(\mathbf{C}_1)+H(\mathbf{C}_2)},
\end{equation}
where $\mathbf{C}_1$ and $\mathbf{C}_2$ denote the community partitions of the two functional networks, $H(\mathbf{C})$ is the Shannon entropy, and $I(\mathbf{C}_1,\mathbf{C}_2)$ is the mutual information between the two partitions. In brief, NMI quantifies the similarity between the community structures of two networks, with a larger value of $I_{\mathrm{NMI}}$ indicating higher similarity. In practice, partitions with $I_{\mathrm{NMI}}<0.3$ are regarded as clearly distinguishable~\cite{Newman2004,Karrer2008,Meil2007}.

\vspace{-0.1cm}
\begin{table}[h!]
\caption{\label{tab1} Pairwise values of $I_{\mathrm{NMI}}$ for the functional networks shown in Fig.~\ref{fig3}. Here, $b$ and $\sigma$ denote the bifurcation parameters of the Lorenz and R\"ossler systems, respectively.}
    \vspace{10mm}
    \begin{ruledtabular}
    \begin{tabular}{ccccc}
         & $b=1.4$ & $b=1.38$ & $\sigma=3.4$ & $\sigma=3.286$ \\
        \hline
     $b=1.4$ & n/a & 0.97 & 0.27 & 0.28 \\
     $b=1.38$ & 0.97 & n/a & 0.27 & 0.28 \\
     $\sigma=3.4$ & 0.27 & 0.27 & n/a & 0.96 \\
     $\sigma=3.286$ & 0.28 & 0.28 & 0.96 & n/a \\
    \end{tabular}
    \end{ruledtabular}
\end{table}

Table~\ref{tab1} lists the pairwise values of $I_{\mathrm{NMI}}$ for the functional networks shown in Fig.~\ref{fig3}, from which two distinct features are revealed. First, the value of $I_{\mathrm{NMI}}$ between networks corresponding to different systems is very small. For instance, the $I_{\mathrm{NMI}}$ between the functional network of the Lorenz system with parameter $b=1.4$ and that of the R\"ossler system with parameter $\sigma=3.4$ is about $0.27$, indicating that the reservoir is self-organized into distinct functional configurations when representing different systems. This finding is consistent with the results reported in Refs.~\cite{RC:MFRC2025,RC:LZX2020}. Second, the value of $I_{\mathrm{NMI}}$ between the functional networks of different states from the same system is very large. Specifically, for the Lorenz system, the $I_{\mathrm{NMI}}$ between the two states ($b=1.4$ and $b=1.38$) is about $0.97$; for the R\"ossler system, the $I_{\mathrm{NMI}}$ between the two states ($\sigma=3.4$ and $\sigma=3.286$) is about $0.96$. These large values indicate that, when representing different states of the same system, essentially the same set of nodes participates in the reservoir in a coordinated manner. In other words, the functional network is not sensitive to variations in the system parameter. This finding extends the results reported in Refs.~\cite{RC:MFRC2025,RC:LZX2020}, where the analysis was restricted to the sampled states. Since the machine is operated in the closed-loop mode when replicating the attractors, these distinct features of $I_{\mathrm{NMI}}$ also highlight the different roles played by the parameter-control and system-label channels in modulating the reservoir dynamics: the label vector switches the reservoir dynamics between two distinct patterns, whereas the bifurcation parameter only slightly modifies the dynamical pattern.

Returning to the questions posed earlier in this section, our preliminary conclusions from the above analysis are as follows. First, the distinguishability of the functional networks corresponding to different systems is slightly lower in the dual-channel RC scheme than in the MFRC scheme reported in Ref.~\cite{RC:MFRC2025} (where $I_{\mathrm{NMI}}\approx0.02$ for the chaotic Lorenz and R\"ossler systems). Second, for the same system, the functional networks of the sampled and unseen states are similar. These results are not unexpected, since the proposed scheme is designed to infer not only the dynamics of the sampled states, but also unseen dynamics not included in the training set, i.e., to reconstruct the bifurcation diagrams of both systems from partial observations. As a result, the machine must balance accuracy and adaptability in learning the system dynamics, which makes the $I_{\mathrm{NMI}}$ between the functional networks of different systems slightly larger than that in MFRC~\cite{RC:MFRC2025}. (Similar behavior is also observed for other sampled states. Please see the Supplementary Material for details.)

\section{Experimental Results}
\label{sec:experiment}

\begin{figure*}[tbp]
\begin{center}
\includegraphics[width=\textwidth]{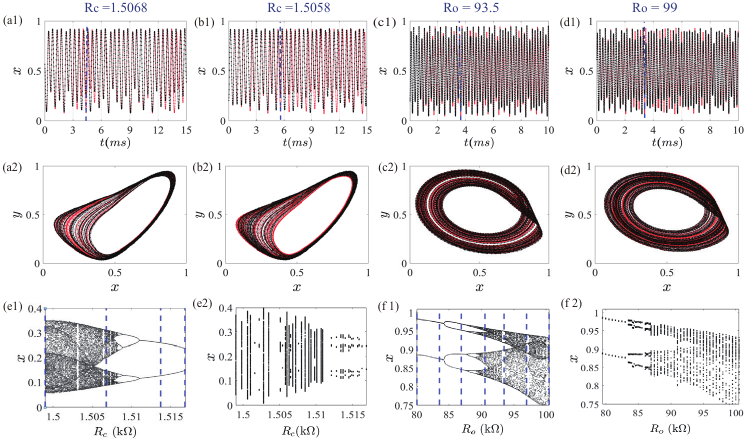}
\centering
\caption{Inferring the dynamics of the Chua and R\"ossler circuits by dual-channel RC. (a,b,e) Results for the Chua circuit. (a1,a2) State evolution and strange attractor for the sampled state $R_c=1.5068$. (b1,b2) State evolution and strange attractor for the unseen state $R_c=1.5058$. (e1) Bifurcation diagram inferred by the machine. (e2) Ground-truth bifurcation diagram obtained from experiments. (c,d,f) Results for the R\"ossler circuit. (c1,c2) State evolution and strange attractor for the sampled state $R_o=93.5$. (d1,d2) State evolution and strange attractor for the unseen state $R_o=99$. (f1) Bifurcation diagram inferred by the machine. (f2) Ground-truth bifurcation diagram obtained from experiments. In (a1)--(d2), black curves denote the experimental results and red curves denote the results inferred by the machine; the vertical dashed lines mark the prediction horizons. In (e1) and (f1), the vertical dashed lines indicate the sampled states from which the training data are collected.}
\label{fig4}
\end{center}
\end{figure*}

We next investigate the performance of the proposed scheme on experimental data acquired from two electrical circuits with distinct dynamics: the Chua circuit~\cite{RC:LHB2024} and the R\"ossler circuit~\cite{Hu_gang:rossler}. Unlike the data generated by numerical simulation in model systems, experimental data are corrupted by noise, which poses an additional challenge for learning the system dynamics. In our experiments, the dynamics of the Chua circuit are modulated by a resistor, which is varied within the range $R_c\in[1.4990,1.5169]$ (k$\Omega$); the dynamics of the R\"ossler circuit are likewise modulated by a resistor, which is varied within the range $R_o\in[80,100.5]$ (k$\Omega$). For the Chua circuit, the sampled states are chosen at $R_c\in\{1.4990,1.5068,1.5137,1.5169\}$. The circuit dynamics are chaotic for $R_c=1.4990$ and $1.5068$, and periodic for $R_c=1.5137$ and $1.5169$. For the R\"ossler circuit, the sampled states are chosen at $R_o\in\{80, 83.5, 86.9, 90.5, 93.5, 97,100.5\}$. The circuit dynamics are periodic for $R_o\in\{80, 83.5, 86.9\}$ and chaotic for $R_o\in\{90.5, 93.5, 97, 100.5\}$. For each sampled state, a time series containing $2\times10^{4}$ data points is recorded for the Chua circuit, and a time series containing $5\times10^{4}$ data points is recorded for the R\"ossler circuit. All data are normalized to the range $[0,1]$ (please see the Supplementary Material for more details about the experimental setups and data acquisition).

The training and validation procedures are the same as those used for the model systems. As before, we label the Chua and R\"ossler circuits by the vectors $[1,0]^T$ and $[0,1]^T$, respectively. The time evolution of the $x$ variable of Chua's circuit predicted by the machine for the sampled state $R_c=1.5068$ is shown in Fig.~\ref{fig4}(a1). We see that the state evolution is accurately predicted for about $4.5~\mathrm{ms}$ ($10$ oscillations). The long-term evolution of the system predicted by the machine is shown in Fig.~\ref{fig4}(a2), where the attractor inferred by the machine agrees well with the ground truth. The results for an unseen state of Chua's circuit with $R_c=1.5058$, which is not included in the training phase, are shown in Figs.~\ref{fig4}(b1) and (b2). We see that, for this unseen state, the prediction horizon is about $6~\mathrm{ms}$ ($12$ oscillations), and the inferred attractor also agrees well with the ground truth. By increasing $R_c$ from $1.4990$ to $1.5169$ with an increment of $\Delta R_c=1\times10^{-4}$ in the parameter-control channel, we reconstruct the bifurcation diagram of Chua's circuit from the machine outputs, as shown in Fig.~\ref{fig4}(e1). The bifurcation diagram is plotted using the local minimum of the $x$ variable. It is seen from Fig.~\ref{fig4}(e1) that, as $R_c$ increases, the system undergoes rich bifurcations, including chaotic regimes, periodic windows, and an inverse period-doubling scenario. For comparison, the ground-truth bifurcation diagram obtained from experiments is shown in Fig.~\ref{fig4}(e2). In obtaining the experimental bifurcation diagram, a total of $80$ experiments are conducted, with $R_c$ ranging from $1.4990$ to $1.5169$ with an increment of $\Delta R_c=2\times10^{-4}$. Compared with Fig.~\ref{fig4}(e2), the bifurcation diagram inferred by the machine in Fig.~\ref{fig4}(e1) is of higher quality and exhibits a cleaner and clearer structure.

Using the same machine, we next infer the dynamics of the R\"ossler circuit by inputting $[0,1]^T$ into the system-label channel. The results for the sampled state with parameter $R_o=93.5$ are shown in Figs.~\ref{fig4}(c1) and (c2). We see that the prediction horizon is about $4~\mathrm{ms}$ ($19$ oscillations), and the inferred attractor agrees well with the ground truth. Similar results are obtained for the unseen state with parameter $R_o=99$, which is not included in the training set, as shown in Figs.~\ref{fig4}(d1) and (d2). By varying $R_o$ within the range $[80,100.5]$ with an increment of $\Delta R_o=0.2$, we reconstruct the bifurcation diagram from the machine outputs, as shown in Fig.~\ref{fig4}(f1). The experimental bifurcation diagram is shown in Fig.~\ref{fig4}(f2). We see that, compared with the experimental results, the bifurcation diagram inferred by the machine reveals more details of the bifurcation structure.

Additional analysis has been conducted to benchmark the proposed scheme against the conventional PARC scheme in reconstructing the bifurcation diagrams of the circuits. The same set of sampled states is used in this comparison. For the conventional PARC scheme~\cite{KLW:2021,RC:FHW2021,RC:HBL2025}, two separate machines are trained, each for inferring one circuit, whereas for the proposed scheme, a single machine is trained to infer both circuits. The results show that the bifurcation diagrams reconstructed by the proposed scheme are consistent with those obtained using the conventional PARC scheme (please see the Supplementary Material for details).

\section{Discussion and Conclusion}

Motivated by recent studies on multifunctional machines~\cite{RC:LZX2020,RC:Flynn2021,KLW2024,RC:MFRC2025}, we have proposed a new RC scheme capable of inferring the dynamics of two distinct chaotic systems. Compared with existing MFRC schemes, the advantages of the proposed scheme are mainly reflected in the following two aspects. First, the dynamics of the target systems can be reproduced more accurately and retrieved more reliably. Two major challenges currently encountered in multifunctional machines are that (1) the retrieval of stored attractors is not reliable, i.e., depending on the initial condition of the reservoir network, the same input label may lead to different outputs, and (2) the retrieved dynamics are seriously distorted, i.e., the attractors inferred by the machine do not closely resemble the ground truth. Both challenges are effectively addressed by the proposed scheme, as demonstrated in Figs.~\ref{fig2}(a)--(d) for the model systems and Figs.~\ref{fig4}(a)--(d) for the electrical circuits. Second, the proposed machine is capable of inferring unseen dynamics for both systems. Existing multifunctional machines, while able to replicate multiple systems with distinct dynamics, cannot infer dynamics that are not included in the training phase. Here, by incorporating a parameter-control channel into the input layer, as in PARC~\cite{KLW:2021,RC:Kim2021,RC:FHW2021,RC:LHB2024}, the proposed scheme is able not only to replicate the dynamics of the sampled states, but also to infer unseen dynamics for each individual system, as demonstrated in Figs.~\ref{fig2}(e) and (f) for the model systems and Figs.~\ref{fig4}(e) and (f) for the electrical circuits. Taken together, the proposed scheme combines the advantages of multifunctional machines and PARC, enabling the inference of the bifurcation diagrams of two distinct systems from partial observations.

While our results demonstrate the capability of the proposed scheme in learning different dynamics, the present study is still preliminary, and many important questions remain to be addressed. First, for the purpose of demonstration, we have considered only the simplest case of learning two distinct systems. Although it is generally expected that the machine performance will degrade as the number of target systems increases~\cite{KLW2024,RC:MFRC2025}, a detailed analysis is still needed to quantify the impact of this increase on the accuracy and reliability of the replicated attractors, as well as on the quality of the reconstructed bifurcation diagrams. Second, for simplicity, we have focused on chaotic systems with the same state-space dimension in demonstrating the application of the dual-channel scheme. Extending the proposed scheme to dynamical systems of different dimensions is another important issue that remains to be explored. One possible approach is to adopt the padding technique in data processing~\cite{RC:Versatile2025}, yet the feasibility of this strategy remains to be investigated. Third, although the proposed machine is able to reconstruct the bifurcation diagrams of two distinct systems from the time series of only a small number of sampled states, a detailed analysis of the reconstruction accuracy is still lacking. So far, we have shown only that the reconstructed diagrams capture the main features of the ground truth, such as the transition from periodic to chaotic dynamics and the periodic windows embedded in the chaotic regime. A quantitative characterization of the accuracy and precision of the reconstructed bifurcation diagrams is still lacking. Finally, the working mechanism of the proposed machine is still not fully understood. Although our analysis shows that distinct systems are represented by distinct functional networks, it remains unclear how these functional networks are modulated by the bifurcation parameters. Since functional networks characterize the collective dynamics of the reservoir, it is natural to expect that they may also exhibit distinct bifurcation behavior. Therefore, an in-depth analysis of the bifurcation behavior of the reservoir network is crucial for a more complete understanding of the working mechanism of the proposed machine.

To summarize, we have proposed a new RC scheme that enables a single machine to infer the dynamics of two distinct systems. By combining the ideas of MFRC through the incorporation of a system-label channel into the input layer and PARC through the incorporation of a parameter-control channel, the proposed machine is able to infer not only the dynamics of the sampled states, but also unseen dynamics of both systems that are not included in the training process. More importantly, using the proposed scheme, we have successfully reconstructed the bifurcation diagrams of the classical Lorenz and R\"ossler systems in simulations, as well as those of the Chua and R\"ossler circuits in experiments. Analysis based on functional networks suggests that the reservoir responds in markedly different ways when representing different systems. These findings extend our understanding of the learning capability of RC and provide practical insights into the design of high-performance multifunctional machines.

\vspace{0.5cm}
\begin{acknowledgments}
This work was supported by the National Natural Science Foundation of China (NNSFC) under Grant No.~12275165 and No.~42461144209. XGW acknowledges support from the Shaanxi Fundamental Science Research Project for Mathematics and Physics under Grant No.~25JSZ009.
\end{acknowledgments}

\vspace{-0.5cm}
\section*{Data availability}
The data and code supporting the findings of this study are openly available in Ref.~\cite{Github}.

\end{document}